\documentclass[a4paper]{article}

\usepackage{geometry}
\usepackage[figuresright]{rotating}
\usepackage{graphicx}
\usepackage{longtable}
\usepackage{array}
\usepackage{booktabs}

\usepackage{multirow}
\usepackage{makecell}

\usepackage{graphicx}
\usepackage{tensor}
\usepackage{mathtools}
\usepackage{amsmath}
\allowdisplaybreaks[4]
\usepackage{upgreek}

\usepackage{bm}

\usepackage[noblocks]{authblk}

\usepackage[numbers,sort&compress]{natbib}

\usepackage{CJK}
\usepackage{txfonts}
\usepackage{times}
\usepackage{ragged2e}
\usepackage{indentfirst}

\usepackage{verbatim}

\usepackage{ntheorem}

\usepackage{color}

\renewcommand{\raggedright}{\leftskip=0pt \rightskip=0pt plus 0cm}

\begin{document}

\title{Two-party quantum private comparison based on \\ eight-qubit entangled state}
\author{Peiru Fan$^{1}$, Atta Ur Rahman$^{3}$, Zhaoxu Ji$^{3,*}$,
Xiangmin Ji$^{2,^\dag}$, Zhiqiang Hao$^{1}$, Huanguo Zhang$^{3,\ddag}$
\\
{\small {
$^1$China Industrial Control Systems Cyber Emergency Response Team, Beijing 100040, China	\\
$^2$College of Computer Information Science, Fujian Agriculture and Forestry University, Fuzhou 35002,China \\
$^3$Key Laboratory of Aerospace Information Security and Trusted Computing,
Ministry of Education, School of Cyber Science and Engineering,
Wuhan University, Wuhan 430072 China \\
$^*$jizhaoxu@whu.edu.cn, $^\dag$jixm168@126.com, $^{\ddag}$liss@whu.edu.cn}}}
\date{}


\maketitle

\begin{abstract}

The purpose of quantum private comparison (QPC) is to solve ``Tierc\'e problem'' using quantum mechanics laws,
where the ``Tierc\'e problem'' is to judge whether the secret data of two participants are equal 
under the condition of protecting data privacy.
Here we consider for the fist time the usefulness of eight-qubit entangled states for QPC by proposing a new protocol.
The proposed protocol only adopts necessary quantum technologies such as
preparing quantum states and quantum measurements without using
any other quantum technologies (e.g., unitary operations and entanglement swapping),
thus the protocol have advantages in quantum device consumption.
The measurements adopted only include single-particle measurements,
which is easier to implement than entangled-state measurements under the existing technical conditions.
The proposed protocol takes advantage of the entanglement characteristics of the eight-qubit entangled state,
and uses joint computation, decoy photon technology, 
the keys generated by a quantum key distribution protocol to ensure data privacy.
We show that when all single-particle measurements in the proposed protocol are replaced
by Bell measurements, the purpose of the protocol can also be achieved. We also show that the proposed protocol 
can be changed into a semi-quantum protocol with a few small changes.

\end{abstract}


\noindent
\textbf{PACS number(s)}: 03.65.Ud, 03.67.Dd



\rmfamily
\fontsize{10}{13}\selectfont

\raggedright{\section{Introduction}}

Cryptography includes various technologies of encryption, decryption, authentication and signature etc.,
which is the core technology to ensure information security \cite{ZhangHG58112015}.
Unfortunately, cryptography is facing severe challenges from 
quantum information technology \cite{ZhangHG162019}.
In this context, quantum cryptography (QC) has been widely concerned because some quantum mechanics principles
such as Heisenbergs uncertainty principle and quantum noncloning theorem provide unconditional security for it
\cite{ZhangHG162019}.
In the last two decades, a lot of remarkable progress has been made in the experiment of QC,
and began the industrialization development \cite{ZhangHG162019}.
QC is likely to develop into an important force for the next generation of technologies
in cryptography \cite{ZhangHG162019}.

As an important branch of QC, quantum secure multi-party computation (QSMC) \cite{ZhangHG162019}
is the counterpart of the classical secure multi-party computation \cite{YaoAC}
in the realm of quantum mechanics. Since Bennett and Brassard proposed the pioneering work of QC \cite{Bennett},
a large number of  QSMC protocols have been proposed,
such as quantum private query \cite{FGao6272019}, quantum secret sharing \cite{HilleryM,TsaiC34272019},
quantum anonymous voting and surveying \cite{VaccaroJA,JiZX182019},
and quantum oblivious set-member decision \cite{ShiRH922019}.
In 2009, Yang et al. proposed quantum private comparison (QPC),
which also belongs to the field of QSMC \cite{YangYG4252009}.
QPC allows two participants (commonly known as Alice and Bob)
who do not trust each other to judge whether the private data they have are equal,
and uses the principles of quantum mechanics to protect data privacy
\cite{ZhangHG162019,SunZW5212013,JiZX962021,JiZX4592020,JiZX5852022,LangYF57102018}.
Its importance lies in the fact that it can solve ``Tierc\'e problem'' \cite{FBoudot1112001} and has potential applications in
private bidding and auctions, authentication, identification, and secret ballot elections \cite{YangYG4252009}.
At present, various multi-particle entangled states, such as GHZ states \cite{ChenXB28372010,LiuW51112012},
W states \cite{LiJ5372014,LiuW2842011}, highly entangled six-qubit genuine states \cite{JiZX6562016}
and $\upchi$-type states \cite{LiuW5162012}, have been considered to design QPC protocols.

Recently, we investigated the utility of the maximally entangled seven-qubit state for QPC \cite{JiZX34282019} 
by proposing a QPC protocol, where the entanglement correlation of this state
is used to complete the private comparison and protect data privacy.
However, one problem is ignored in this protocol and has not yet been solved, that is, 
how to use the keys generated by quantum key distribution (QKD) \cite{ZhangHG162019,Bennett} 
to ensure the security of joint computation between two participants when the two participants are in different places.
We consider in this paper for the first time the usefulness of the eight-qubit entangled state
for comparing two participants' data by proposing a new QPC protocol, 
and address the neglected problem.
In the proposed protocol, a semi-honest third party (conventionally called TP in QPC) is introduced,
who is curious about the two participants' data but honestly assists them to complete the protocol process 
without colluding with any one of them.
We also show two variants of the proposed protocol: one is to replace the single-particle measurements measurements 
in the protocol with Bell measurements, and the other is to transform the proposed protocol into a semi-quantum protocol 
by slightly changing the decoy photon technology used for security checking and QKD used to generate keys 
(see Sec. \ref{variants}).

The rest of this paper is arranged as follows. In Sec. 2, we provide a brief 
review of our previously proposed protocol presented in Ref. \cite{JiZX34282019}.
Sec. 3 introduces the eight-qubit entangled state
used as information carriers and the prerequisites of the proposed protocol.
Then the steps of the proposed protocol are described in detail.
Sec. 4 provides some useful discussions. Sec. 5 analyzes the protocol security,
including the external and internal attacks. 
In Sec. 6, we show that by making some small changes, 
the proposed protocol can be transformed into a Bell-measurement-based protocol and a semi-quantum protocol respectively.
The summary of this paper is made in the last section.


\section{Brief review on our previous work}

Before presenting the new QPC protocol, we would like to briefly review the process of our previously proposed protocol,
which is shown in Ref. \cite{JiZX34282019}.
The QPC protocol presented in Ref. \cite{JiZX34282019} is based on the maximally entangled seven-qubit state,
which can be expressed by the following two forms \cite{JiZX34282019}:
\begin{align}
\left|\Psi\left(p^1,p^2,\dots,p^7\right)\right\rangle_{1234567}
=\frac 1{\sqrt{32}} (
& \left|0000000\right\rangle+\left|0000011\right\rangle
+\left|0001101\right\rangle + \left|0001110\right\rangle + \left|0010001\right\rangle  	\notag \\
-&\left|0010010\right\rangle + \left|0011100\right\rangle - \left|0011111\right\rangle 	
-\left|0100101\right\rangle - \left|0100110\right\rangle + \left|0101000\right\rangle 	\notag \\
+&\left|0101011\right\rangle + \left|0110100\right\rangle - \left|0110111\right\rangle 
-\left|0111001\right\rangle + \left|0111010\right\rangle - \left|1000100\right\rangle 		\notag \\
-&\left|1000111\right\rangle + \left|1001001\right\rangle + \left|1001010\right\rangle 	
+\left|1010101\right\rangle - \left|1010110\right\rangle - \left|1011000\right\rangle		\notag \\
+&\left|1011011\right\rangle + \left|1100001\right\rangle + \left|1100010\right\rangle 
+\left|1101100\right\rangle + \left|1101111\right\rangle + \left|1110000\right\rangle 	\notag \\
-&\left|1110011\right\rangle + \left|1111101\right\rangle -
\left|1111110\right\rangle	)_{1234567} \notag \\
=  \frac 1 4  \Big\{\big[(	
& \left|0000\right\rangle + \left|0101\right\rangle) \left|0\right\rangle +
( \left|1101\right\rangle - \left|1000\right\rangle) \left|1\right\rangle	\big]
\left|\phi^{+}\right\rangle      							
\notag \\
+  \big[(
& \left|1001\right\rangle + \left|1100\right\rangle ) \left|0\right\rangle +
(\left|0001\right\rangle - \left|0100\right\rangle) \left|1\right\rangle	\big]
\left|\psi^{+}\right\rangle
\notag \\
+  \big[(	
& \left|0010\right\rangle - \left|0111\right\rangle) \left|0\right\rangle +
(\left|1010\right\rangle + \left|1111\right\rangle) \left|1\right\rangle	\big]
\left|\psi^{-}\right\rangle
\notag \\
+ \big[(	
&\left|1110\right\rangle - \left|1011\right\rangle) \left|0\right\rangle +
(\left|0011\right\rangle + \left|0110\right\rangle) \left|1\right\rangle	\big]
\left|\phi^{-}\right\rangle	\Big\}_{1234567},
\end{align}
where the symbols $p^1,p^2,\dots,p^7$ represent the seven particles,
the numbers $1, 2,\dots,7$ indicate their orders,
and $\left|\phi^{\pm}\right\rangle$ and $\left|\psi^{\pm}\right\rangle$ are the well-known Bell states
\begin{align}
\label{Bell-sate}
\left|\phi^{\pm}\right\rangle = \frac 1{\sqrt{2}}
\Big(   \left|00\right\rangle \pm \left|11\right\rangle  \Big),
\phantom{i}
\left|\psi^{\pm}\right\rangle = \frac 1{\sqrt{2}}
\Big(   \left|01\right\rangle \pm \left|10\right\rangle  \Big).
\end{align}
The participants of the protocol perform single-particle measurements on the first five particles 
and Bell measurements on the last two particles. Then they encode the measurement results of the first four particles as follows,
\begin{align}
\begin{cases}
\left|0\right\rangle \leftrightarrow 0,
\\
\left|1\right\rangle \leftrightarrow 1.
\end{cases}
\end{align}
and encode the measurement results of the last three particles as follows,
\begin{align}
\label{encoding-rule}
\begin{cases}
\left|0\right\rangle \& \left|\phi^+\right\rangle \leftrightarrow 00,	\phantom{i}
\left|1\right\rangle \& \left|\phi^+\right\rangle \leftrightarrow 10,	\phantom{i}
\left|1\right\rangle \& \left|\psi^+\right\rangle \leftrightarrow 01,	\phantom{i}
\left|0\right\rangle \& \left|\psi^+\right\rangle \leftrightarrow 11,	
\\
\left|0\right\rangle \& \left|\psi^-\right\rangle \leftrightarrow 10, 	\phantom{i}
\left|1\right\rangle \& \left|\psi^-\right\rangle \leftrightarrow 00, 	\phantom{i}
\left|1\right\rangle \& \left|\phi^-\right\rangle \leftrightarrow 11, 	\phantom{i}
\left|0\right\rangle \& \left|\phi^-\right\rangle \leftrightarrow 01.
\end{cases}
\end{align}
Let us denote the classical bits corresponding to the measurement results of the first four particles as
$b_1,b_2,b_3,b_4$ respectively,
and the ones corresponding to the measurement results of the last three particles as $b_{567}$.
From the entanglement correlation of the maximally entangled seven-qubit state, one can get
\begin{align}
\label{correlation}
b_1 b_2 \oplus b_3 b_4 = b_{567}.
\end{align}

Two participants who do not trust each other, Alice and Bob, have the secret data $X$ and $Y$ respectively,
where the binary representations of $X$ and $Y$ are given by
\begin{align}
\begin{cases}
X = \left( x_{1}, x_{2},\dots,x_{n} \right), \phantom{i} x_{k} \in \{ 0, 1 \} \phantom{i} \forall k=1, 2,\dots, n,
\\
Y = \left( y_{1}, y_{2},\dots,y_{n} \right), \phantom{i} y_{k} \in \{ 0, 1 \} \phantom{i} \forall k=1, 2,\dots, n,
\end{cases}
\end{align}
such that
\begin{align}
\begin{cases}
X=\sum_{k=1}^{n} x_{k} \cdot 2^{k-1}, \\
Y=\sum_{k=1}^{n} y_{k} \cdot 2^{k-1}, \\
2^{n} \le \max \{ X,Y \} < 2^{n+1},
\end{cases}
\end{align}
Alice and Bob want to judge whether $X=Y$ with the help of a third party TP
who is curious about their data but will not collude with them. 
We can describe this in the form of secure multiparty computing as follows:
\begin{align}
f(X,Y) =
\begin{cases}
1, \phantom{i} X=Y;
\\
0, \phantom{i} X \ne Y.
\end{cases}
\end{align}
That is, when the input data $X$ is equal to $Y$, the output of function $f(X,Y)$ is 1,
such that TP tells Alice and Bob that their data is the same publicly, and vice versa.
In the initial stage of the protocol, Alice and Bob divide the binary representations of their data into the following 
$\lceil N/2 \rceil$ groups:
\begin{align}
\begin{cases}
G_a^1, G_a^2,\ldots, G_a^{\lceil \frac N 2 \rceil},
\\
G_b^1, G_b^2,\ldots, G_b^{\lceil \frac N 2 \rceil},
\end{cases}
\end{align}
where $ G_a^i (G_b^i) $ includes two classical bits.
TP generates $\lceil N/2 \rceil$ copies of $\left|\Psi\left(p^1,p^2,\dots,p^7\right)\right\rangle_{1234567}$,
and take out the particles $p^1,p^2,p^3,p^4$ from each state, 
and then send them to Alice and Bob using decoy photon technology \cite{JiZX34282019}.
Subsequently, Alice and Bob measure the received particles respectively,
and mark the classical bits corresponding to the measurement results by
\begin{align}
\begin{cases}
C_a^1, C_a^2,\ldots, C_a^{\lceil \frac N 2 \rceil},
\\
C_b^1, C_b^2,\ldots, C_b^{\lceil \frac N 2 \rceil}.
\end{cases}
\end{align}
$\forall i = 1,2,\dots,\lceil \frac N 2 \rceil$, they calculate $C_a^i \oplus G_a^i$ and $C_b^i \oplus G_b^i$, respectively.
Then they cooperate together to calculate
\begin{align}
R_{ab}^i = C_a^i \oplus G_a^i \oplus C_b^i \oplus G_b^i,
\end{align}
and announce the calculation results to TP. In the final stage of the protocol,
TP performs single-particle measurements on the particles $p^5$
and Bell measurements on the particles $p^6,p^7$ in each state.
According to the encoding rule shown in Eq. \ref{encoding-rule}, TP denote the classical bits
corresponding to his measurement results as $C_c^i$.
Finally, he computes $R_{ab}^i \oplus C_c^i$,
and tells Alice and Bob that their data are the same iff 
$R_{ab}^i \oplus C_c^i = 00 \phantom{i} \forall i = 1,2,\dots,\lceil \frac N 2 \rceil$.
Otherwise TP tells them that their data are different.

The correctness of the protocol output is easy to verify.
From Eq. \ref{correlation}, one can get $C_a^i \oplus C_b^i = C_c^i$, thus
\begin{align}
R_{ab}^i = R_{ab}^i \oplus C_c^i = C_a^i \oplus G_a^i \oplus C_b^i \oplus G_b^i \oplus C_c^i = G_a^i \oplus G_b^i.
\end{align}
It can be seen that the comparison results published by TP are correct at the end of the protocol.


\section{Proposed QPC protocol}

In what follows, we will first introduce the eight-qubit entangled state used as information carriers, 
and then give the prerequisites of the proposed protocol. We will finally describe the protocol steps.

\subsection{Information carriers}

The form of the eight-qubit entangled state \cite{SadeghiZadehMS5672017} is given by 
\begin{align}
\label{information-carrier}
\left|\varUpsilon\left(p^1,p^2,\dots,p^8\right)\right\rangle
&=\frac 1 4	\big(
\left|00000000\right\rangle+\left|00010001\right\rangle +\left|00100010\right\rangle + \left|00110011\right\rangle  \notag \\
&\quad\phantom{(}+
\left|01000100\right\rangle + \left|01010101\right\rangle + \left|01100110\right\rangle + \left|01110111\right\rangle \notag \\
&\quad\phantom{(}+
\left|10001000\right\rangle + \left|10011001\right\rangle + \left|10101010\right\rangle +\left|10111011\right\rangle \notag \\
&\quad\phantom{(}+
\left|11001100\right\rangle + \left|11011101\right\rangle + \left|11101110\right\rangle + \left|11111111\right\rangle \big)_{12345678},
\end{align}
where the symbols $p^1,p^2,\dots,p^8$ represent the eight particles
and the numbers $1, 2,\dots,8$ indicate their order in a state.

\subsection{Prerequisites}

The prerequisites of the proposed protocol are given by

\begin{enumerate}

\item Assume that Alice and Bob have the private data $X$ and $Y$ respectively,
and that $x_{1},x_{2},\ldots,x_{N}$ and $y_{1}, y_{2},\ldots,y_{N}$ 
($ x_{j}, y_{j} \in \{ 0, 1 \} \phantom{1} \forall j=1,2,\ldots,N$) are
the binary representations of $X$ and $Y$ respectively (i.e., $X=\sum_{j=1}^{N} x_j2^{j-1},Y=\sum_{j=1}^{N} y_j2^{j-1}$).
Alice and Bob want to know whether $X = Y$ with the assistance of a semi-honest third party (TP) who
is curious about their data but will not collude with them.

\item Alice and Bob are in the same place, which means that they do not need to
use a channel to complete the communication between them.

\item Alice (Bob) divides $x_{1},x_{2},\ldots,x_{N} (y_{1}, y_{2},\ldots,y_{N})$ into $\lceil N/2 \rceil$ groups in turn,
and marks them by
\begin{equation}
G_a^1, G_a^2,\ldots, G_a^{ \lceil \frac N 2 \rceil } 
\left( G_b^1, G_b^2,\ldots, G_b^{\lceil \frac N 2 \rceil}	\right),
\end{equation}
where $ G_a^i (G_b^i) $ includes two classical bits, and $i = 1,2,\ldots,\lceil N/2 \rceil$ throughout this paper.
Note here that when $ N $ mod $ 2 = 1 $, one ``0'' should be added into the last group
$ G_a^{ \lceil N/2 \rceil}$ $( G_b^{\lceil N/2  \rceil} ) $.

\item TP, Alice and Bob agree on the two public coding rules:
$\left|0\right\rangle \leftrightarrow 0$ and $\left|1\right\rangle \leftrightarrow 1$.

\end{enumerate}

\subsection{Protocol steps}

\begin{figure}[h]
\centerline{\includegraphics[width=5.5in]{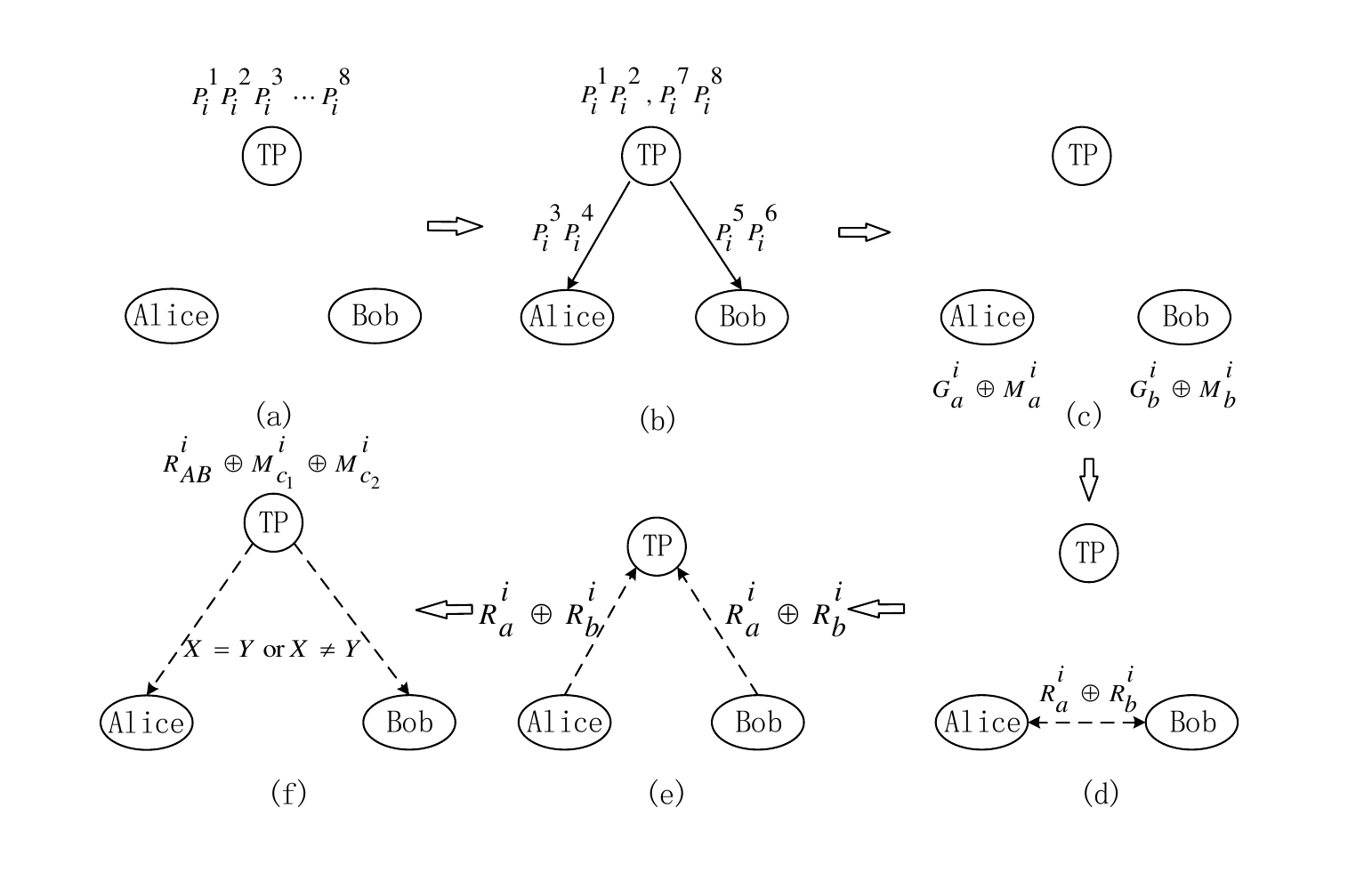}}
\vspace*{8pt}
\caption{Simple flow chart describing the process of the proposed protocol .
The the dashed arrow lines denote classical channels, and solid arrow ones represent quantum channels.
Note here that the eavesdropping checking and external eavesdroppers are omitted.\protect\label{fig1}}
\end{figure}

The steps of the proposed protocol (the corresponding flow chart is shown in figure \ref{fig1}) are given by

\begin{enumerate}

\item
TP prepares $\lceil \frac N 2 \rceil$ copies of $\left|\varUpsilon\left(p^1,p^2,\dots,p^8\right)\right\rangle$
and marks them by
\begin{equation}
\left|\varUpsilon\left(p_1^1,p_1^2,\dots,p_1^8\right)\right\rangle,
\left|\varUpsilon\left(p_2^1,p_2^2,\dots,p_2^8\right)\right\rangle,
\dots,
\left|\varUpsilon\left(p_{\lceil N/2 \rceil}^1,p_{\lceil N/2 \rceil}^2,\dots,p_{\lceil N/2 \rceil}^8\right)\right\rangle,
\end{equation}
where the superscripts $1,2,\dots,8$ represent eight particles in one state.
Then he takes the particles marked by ($p_i^3,p_i^4$) out from $\left|\varUpsilon\left(p_i^1,p_i^2,\dots,p_i^8\right)\right\rangle$
to form a new ordered particle sequence
\begin{equation}
p_1^3,p_1^4,p_2^3,p_2^4,\dots,p_{\lceil N/2 \rceil}^3,p_{\lceil N/2 \rceil}^4,
\end{equation}
marked by $S_a$.
Similarly, he takes the particles marked by ($p_i^5,p_i^6$) out from $\left|\varUpsilon\left(p_i^1,p_i^2,\dots,p_i^8\right)\right\rangle$
to form another sequence
\begin{equation}
p_1^5,p_1^6,p_2^5,p_2^6,\dots,p_{\lceil N/2 \rceil}^5,p_{\lceil N/2 \rceil}^6,
\end{equation}
marked by $S_b$.
The remaining particles in $\left|\varUpsilon\left(p_i^1,p_i^2,\dots,p_i^8\right)\right\rangle$ form the sequence
\begin{equation}
p_1^1,p_1^2,p_1^7,p_1^8,p_2^1,p_2^2,p_2^7,p_2^8,\dots,
p_{\lceil N/2 \rceil}^1,p_{\lceil N/2 \rceil}^2,p_{\lceil N/2 \rceil}^7,p_{\lceil N/2 \rceil}^8,
\end{equation}
denoted as $S_c$.

\item
TP prepares two sequences of decoy photons, marked by $D_a$ and $D_b$ respectively, 
in which each photon is randomly in one of
$ \left|0\right\rangle, \left|1\right\rangle, \left|+\right\rangle, \left|-\right\rangle
[\left|\pm\right\rangle= 1/\sqrt{2} \left( \left|0\right\rangle \pm \left|1\right\rangle  \right)]$.
Then he randomly inserts these photon into $S_a (S_b)$.
Marking the new generated sequences by $ S_a^* (S_b^*)$,
TP sends $ S_a^* $ to Alice and $ S_b^* $ to Bob.

\item
TP and Alice (TP and Bob) use the decoy photons in $D_a (D_b)$ to complete eavesdropping checking.
Once they find an eavesdropper, they terminate the protocol and start over, otherwise they go on to the next step.

\item
Alice (Bob) performs single-particle measurements on each particle
in $ S_a (S_b) $ using $Z$ basis ($\{\left|0\right\rangle, \left|1\right\rangle\}$).
According to the coding rules (see the third prerequisite of our protocol),
they mark the binary numbers corresponding to the measurement results 
of the two particle $p_i^3,p_i^4 (p_i^5,p_i^6)$ by $ M_a^i (M_b^i) $.
Then Alice (Bob) computes $G_a^i \oplus M_a^i (G_b^i \oplus M_b^i $)
and marks the calculation results by $R_a^i (R_b^i)$,
where the symbol $\oplus$ represents the module 2 operation throughout this paper.
Subsequently, Alice and Bob cooperate together to compute $ R_a^i \oplus R_b^i $. 
Marking the computing results by $R_{AB}^i$,  Alice and Bob publicly announce $ R_{AB}^i $ to TP.

\item
After receiving $ R_{AB}^i $, TP measures each particle in $S_c$ by $Z$ basis.
Then TP denotes the measurement outcomes of the particles marked by $p_i^1,p_i^2$ as $M_{c_1}^i$,
and the measurement outcomes of the particles marked by $p_i^7,p_i^8$ as $M_{c_2}^i$.
Finally, TP computes $R_{AB}^i \oplus M_{c_1}^i \oplus M_{c_2}^i$, and denotes the computing results as $R_i$.
Iff $ R_i = 00$, TP can conclude that $X = Y$, otherwise $X \ne Y$.
Finally, TP tells Alice and Bob the comparison result in public.

\end{enumerate}

\section{Discussion}

We have presented the QPC protocol based on the eight-qubit entangled state.
Indeed, compared with the preparation of single-particle states and Bell states,
it may be difficult to prepare multi-particle entangled states such as the eight-qubit entangled state
with the existing technology. Nevertheless, the study of the usefulness of eight-qubit entangled states
and the entangled states containing more qubits in quantum cryptography has attracted considerable interest
in recent years \cite{JiangS6052021,PengJY5552016,CaoZ6082021,ChoudhuryBS1632019}.
On the one hand, these multi-qubit entangled states provide new ideas and ways for quantum cryptography.
On the other hand, the rapid experimental progress in the preparation of
multi-qubit entangled states increases the possibility of the successful preparation
of the eight-qubit entangled state and the ones containing more qubits \cite{MonzT106132011,WangXL120262018}.
Indeed, despite the experimental progress, many branches of quantum cryptography, including QPC,
are still in the stage of theoretical research \cite{ZhangHG162019}.
Like our previous work shown in Ref. \cite{JiZX34282019}, we only do theoretical research in this paper,
and the experimental analysis is beyond the scope of this paper.

In what follows we would first like to verify the correctness of the algorithm adopted in the proposed protocol
and then consider the situation when the two participants in the protocol are in different places. 
Finally, we compare the protocol with several existing ones.

\subsection{Correctness of the algorithm adopted}

Let us demonstrate that the algorithm adopted in the proposed protocol is correct.
From Eg. \ref{information-carrier}, if performing single-particle measurements on each particle in the eight-particle entangled state,
and denoting the binary numbers corresponding to the outcomes of the measurements as $m_1,m_2,\dots,m_8$ in turn,
then we can always get
\begin{align}
m_1  m_2 \oplus m_7 m_8 = m_3  m_4 \oplus m_5 m_6.
\end{align}
Therefore, in the proposed protocol, we can get
\begin{align}
M_{c_1}^i \oplus M_{c_2}^i = M_a^i \oplus M_b^i,
\end{align}
which can also be derived from table \ref{true_table}.
Therefore, in the final step of the protocol, we have
\begin{align}
\label{output}
R_{AB}^i \oplus M_{c_1}^i \oplus M_{c_2}^i = R_a^i \oplus R_b^i \oplus M_{c_1}^i \oplus M_{c_2}^i
= \left(G_a^i \oplus M_a^i \right) \oplus \left(G_b^i \oplus M_b^i \right) \oplus M_{c_1}^i \oplus M_{c_2}^i
= G_a^i \oplus G_b^i.
\end{align}
It can be seen that the protocol output is correct.

\begin{table}[h]
\centering
\setlength{\tabcolsep}{20pt}

\caption{The truth table}
\begin{tabular}{ccccccccc}

\hline
$M_a^i$  &  $M_b^i$ & $M_{c_1}^i$  & $M_{c_2}^i$ & $M_a^i \oplus M_b^i$ & $M_{c_1}^i \oplus M_{c_2}^i$  \\
\hline

00 & 00 & 00 & 00 & 00 & 00 \\

00 & 01 & 00 & 01 & 01 & 01 \\

00 & 10 & 00 & 10  & 10 & 10 \\

00 & 11 & 00 & 11  & 11 & 11 \\

01 & 00 & 01 & 00 & 01 & 01 \\

01 & 01 & 01 & 01 & 00 & 00 \\

01 & 10 & 01 & 10 & 11 & 11 \\

01 & 11 & 01 & 11  & 10 & 10 \\

10 & 00 & 10 & 00 & 10 & 10 \\

10 & 01 & 10 & 01 & 11 & 11 \\

10 & 10 & 10 & 10 & 00 & 00 \\

10 & 11 & 10 & 11 & 01 & 01 \\

11 & 00 & 11 & 00 & 11 & 11 \\

11 & 01 & 11 & 01 & 10 & 10 \\

11 & 10 & 11 & 10 & 01 & 01 \\

11 & 11 & 11 & 11 & 00 & 00 \\

\hline

\end{tabular}
\label{true_table}
\end{table}


\subsection{Participants are in different places}

Let us now discuss the situation of when Alice and Bob are in different places,
in which case we will only change the fourth and fifth steps of the protocol.
Specifically, in the fourth step, Alice and Bob employ QKD
to prepare the shared key string 
$\left\{K_{ab}^1,K_{ab}^2,\dots, K_{ab}^{\lceil N/2 \rceil}\right\}$, where $K_{ab}^i \in \{00,01,10,11\}$.
Similarly, Alice and TP (Bob and TP) use QKD to prepare the shared key string 
$\left\{K_{ac}^1,K_{ab}^2,\dots, K_{ac}^{\lceil N/2 \rceil}\right\} \left(\left\{K_{bc}^1,K_{bc}^2,\dots, K_{bc}^{\lceil N/2 \rceil}\right\}\right)$,
where $K_{ac}^i, K_{bc}^i \in \{00,01,10,11\}$.
Then Alice (Bob) computes 
$G_a^i \oplus M_a^i \oplus K_{ab}^i \oplus K_{ac}^i \left(G_b^i \oplus M_b^i \oplus K_{ab}^i \oplus K_{bc}^i \right)$,
and denotes the computing results as $R_a^i (R_b^i)$. 
Finally, Alice and Bob tell TP $ R_a^i$ and $ R_b^i $ in public, respectively.
In the fifth step, after receiving $R_a^i$ and $ R_b^i $, TP computes 
$R_a^i \oplus R_b^i \oplus M_{c_1}^i \oplus M_{c_2}^i \oplus K_{ac}^i \oplus K_{bc}^i$,
and denotes the calculation results as $R_i$. Likewise, $X = Y$ iff $ R_i = 00$, otherwise $X \ne Y$.

After the changes, we can still get
\begin{align}
R_i 
&= R_a^i \oplus R_b^i \oplus M_{c_1}^i \oplus M_{c_2}^i \oplus K_{ac}^i \oplus K_{bc}^i \notag \\
&= 	\left( G_a^i \oplus M_a^i \oplus K_{ab}^i \oplus K_{ac}^i \right) \oplus 
	\left(G_b^i \oplus M_b^i \oplus K_{ab}^i \oplus K_{bc}^i \right)
	\oplus M_{c_1}^i \oplus M_{c_2}^i \oplus K_{ac}^i \oplus K_{bc}^i \notag \\
&= 	\left( G_a^i \oplus G_b^i \right) \oplus 
	\left( M_a^i \oplus M_b^i \oplus M_{c_1}^i \oplus M_{c_2}^i \right) \oplus \left(K_{ab}^i \oplus K_{ab}^i \right) 
	\oplus \left( K_{ac}^i \oplus K_{bc}^i \oplus K_{ac}^i \oplus K_{bc}^i \right) \notag \\
&=	G_a^i \oplus G_b^i
\end{align}
It can be seen from the above formula that the protocol output is still correct.

\subsection{Comparison with several available protocols}

Let us compare the proposed QPC protocol with several available ones,
including the protocol presented in Ref. \cite{JiZX34282019} we proposed recently.
Throughout the proposed protocol, as necessary quantum technologies in a QPC protocol
(including the preparation of quantum states and quantum measurement technology),
the protocol only adopts single-particle measurement technology instead of 
entanglement measurement technology used by most existing protocols, which is a main advantage of the protocol.
For example, in addition to the  single-particle measurement technology, 
the protocol in Ref. \cite{JiZX34282019} also adopts Bell measurements, 
which puts forward higher requirements for measurement technology.
Furthermore, the proposed protocol does not use unitary operations and entanglement swapping technology 
adopted by many existing ones, hence it has advantages over these protocols in quantum resource consumption.
Like most existing protocols, the proposed protocol utilizes decoy photons to complete eavesdropping checking,
which is easier to implement than using entangled states for eavesdropping checking.

Table \ref{comparison} and \ref{continuation} show the comparison 
result among the proposed protocol and several available ones.
The qubit efficiency in this table is defined as $\frac ct$, in which $c$ indicates the number of classical bits compared
in a comparison, and $t$ represents the number of consumed particles, regardless of the decoy photons prepared by TP
and those consumed in generating the keys by a QKD protocol.
A eight particle in the proposed protocol is used to compare two classical bits, thus the qubit efficiency is 25\%.
For simplicity and clarity, we use the abbreviation BSM to represent Bell state measurements, 
and the abbreviation SPM single-particle measurements in the two tables.

\begin{table}[h]
\setlength{\tabcolsep}{3pt}

\caption{Comparison among the proposed protocol and several available ones.}
\label{comparison}
\centering
\begin{tabular}{ccccccc}

\hline\noalign{\smallskip}

\makecell{Protocols in references}& Ref. \cite{YangYG4252009} & Ref. \cite{ChenXB28372010} & Ref. \cite{LiuW51112012}
& Ref. \cite{LiJ5372014} & Ref. \cite{LiuW2842011} & Ref. \cite{JiZX6562016} \\

\noalign{\smallskip}\hline\noalign{\smallskip}



\makecell{entanglement swapping} & 
$\times$ & $\times$ & $\surd$ & $\surd$ & $\times$ & $\times$\\

\noalign{\smallskip}\noalign{\smallskip}

\makecell{unitary operations} & 
$\surd$ & $\surd$ & $\times$ & $\times$ & $\surd$ & $\times$ \\

\noalign{\smallskip}\noalign{\smallskip}
\makecell{qubit efficiency} & 
25\% & 33\% & 33\% & 33.3\% & 33\% & 33\% \\

\noalign{\smallskip}\noalign{\smallskip}

\makecell{quantum measurements} 
& BSM
& SPM
& BSM
& BSM
& SPM
& SPM and BSM \\

\noalign{\smallskip}\hline

\end{tabular}
\end{table}

\begin{table}[h]
\setlength{\tabcolsep}{3pt}

\caption{Continuation of table \ref{comparison}.}
\label{continuation}
\centering
\begin{tabular}{cccccc}

\hline\noalign{\smallskip}

Ref. \cite{LiuW5162012} & Ref. \cite{JiZX34282019} & Ref. \cite{WangF59112016} 
& Ref. \cite{YeTY5652017} & Ref. \cite{LiC1852019} & \makecell{Proposed protocol} \\

\noalign{\smallskip}\hline\noalign{\smallskip}



$\surd$ & $\times$ & $\surd$ & $\times$ & $\times$ & $\times$ \\

\noalign{\smallskip}\noalign{\smallskip}

$\times$ & $\times$ & $\surd$ & $\times$ & $\surd$ & $\times$ \\

\noalign{\smallskip}\noalign{\smallskip}

25\% & 29\% & 50\% & 40\% & 50\% & 25\%	\\

\noalign{\smallskip}\noalign{\smallskip}

SPM and BSM
& SPM and BSM
& BSM
& SPM and BSM
& BSM
& SPM	\\

\noalign{\smallskip}\hline

\end{tabular}
\end{table}


\section{Security analysis}

This section analyzes the security of the proposed protocol. We will first show that the external attacks
(i.e., the attacks of the eavesdroppers outside the protocol) are invalid for the protocol, 
and then show that the internal attacks, including the attacks of participants and TP, are also invalid.

\subsection{External attack}

This attack against the proposed protocol refers to an eavesdropper (conventionally called Eve) 
attacking the communication channels between the participants and TP to steal the participants' secret data.
When Alice and Bob are in the same place, the proposed protocol uses joint computing to ensure data privacy; 
When they are in different places, the keys generated by a QKD protocol 
are utilized to encrypt the data to ensure data privacy.
At the same time, decoy photons is utilized for eavesdropping checking,
which has been proved that Eve will be caught when she uses some well-known attack means
such as the entanglement-measurement attack, measurement-resend attack 
and intercept-resend attack \cite{JiZX34282019,JiZX14275}.
In the following sections, we would like to demonstrate how the proposed protocol resists these attacks,
and then we will show the protocol security against several other well-known attacks, 
including photon-number-splitting attack, Trojan horse attack, and man-in-middle attack \cite{ContiM1832016}.

\subsubsection{Intercept-resend attack}

This attack adopted by Eve against the proposed protocol means that
Eve intercepts and captures the particles sent by TP to Alice and Bob in the second protocol step,
and prepares fake particles to replace the intercepted ones.
Then she measures each particle in the eight-qubit entangled states after TP
announces the position and base of each decoy photon prepared by TP to Alice and Bob in the third protocol step,
in which case Eve can get the measurement results $M_a^i$ and $M_b^i$ of Alice and Bob.
However, this attack will not succeed, because when Alice and Bob are in an identical location, 
Eve can only get $G_a^i \oplus G_b^i$ but not $G_a^i$ and $G_b^i$ by decryption after Alice and Bob publish $ R_{AB}^i $  	 to TP.
That is, 
\begin{align}
\left(G_a^i \oplus M_a^i \oplus G_b^i \oplus M_b^i \right) \oplus M_a^i \oplus M_b^i = G_a^i \oplus G_b^i.
\end{align}
When Alice and Bob are in different places, their data are encrypted with the keys generated by a QKD protocol, 
and these keys are unknown to Eve, thus Eve's attack will not succeed.
Furthermore, her attack can be detected with the probability of $1-(3/4)^l$ in the process of eavesdropping checking,
in which $l$ represents the number of decoy photons used in eavesdropping checking.

\subsubsection{Measurement-resend attack}

The attack against the proposed protocol means that 
Eve intercepts and captures the particles that Alice and Bob receive from TP, and directly measures them.
Subsequently, she prepares the same quantum states as the measurement results, and then sends them to Alice and Bob.
However, her attack will fail because she cannot distinguish which of the intercepted particles 
are the particles in the eight-qubit entangled states and which are decoy photons.

\subsubsection{Entanglement-measurement attack}

The form of security proof against entanglement-measurement attacks 
varies with the quantum states utilized in a QSMC protocol. 
Recently, we proved the security of the QPC protocol based on the maximally entangled seven-qubit states
against this attack \cite{JiZX34282019}.
Let us now follow the security proof in Ref. \cite{JiZX34282019} to prove the security against this attack.
Without losing generality, let us consider the case that Eve attacks the quantum channels between Alice and TP.
Let us denote Eve’s unitary operator as $U$, whose properties are given by
\begin{align}
\label{operator}
U \left|0\right\rangle \left|\varepsilon\right\rangle =
\lambda_{00} \left|0\right\rangle \left| \epsilon_{00} \right\rangle + \lambda_{01} \left|1\right\rangle \left| \epsilon_{01} \right\rangle,
U \left|1\right\rangle \left|\varepsilon\right\rangle =
\lambda_{10} \left|0\right\rangle \left| \epsilon_{10} \right\rangle + \lambda_{11} \left|1\right\rangle \left| \epsilon_{11} \right\rangle,
\end{align}
where $\left|\varepsilon\right\rangle$ is an ancillary particle,
$\left| \epsilon_{00} \right\rangle$, $\left| \epsilon_{01} \right\rangle$,
$\left| \epsilon_{10} \right\rangle$, $\left| \epsilon_{11} \right\rangle$
are four pure states determined uniquely by the unitary operator $U$, and
\begin{align}
\label{condition_1}
||\lambda_{00}||^2+||\lambda_{01}||^2 = 1, ||\lambda_{10}||^2+||\lambda_{11}||^2 = 1.
\end{align}
According to the security proof presented in Ref. \cite{JiZX34282019}, 
when the unitary operator $U$ is performed on $\left|+\right\rangle$ and $\left|-\right\rangle$,
if Eve does not want to be detected during eavesdropping checking, the unitary operator $U$ needs to satisfy
\begin{align}
\label{condition_2}
\lambda_{01} = \lambda_{10} = 0,
\lambda_{00} \left| \epsilon_{00} \right\rangle = \lambda_{11} \left| \epsilon_{11} \right\rangle.
\end{align}
From Eqs. \ref{operator}, \ref{condition_1} and \ref{condition_2}, one can get
\begin{align}
\label{condition_3}
\lambda_{00} = \lambda_{11} =1,
\left| \epsilon_{00} \right\rangle = \left| \epsilon_{11} \right\rangle,
U \left|0\right\rangle \left|\varepsilon\right\rangle = \left|0\right\rangle \left|\epsilon_{00}\right\rangle,
U \left|1\right\rangle \left|\varepsilon\right\rangle = \left|1\right\rangle \left|\epsilon_{11}\right\rangle.
\end{align}
When the unitary operator $U$ is performed on the particles in a eight-qubit entangled state,
let us consider the extreme case, that is, $U$ is performed on the particles marked by $p_i^3,p_i^4$.
According to Eq. \ref{condition_3}, one can get
\begin{align}
& I \otimes I \otimes U \otimes U  \otimes I \otimes I \otimes I \otimes I 
\left|\varUpsilon\left(p^1,p^2,\dots,p^8\right)\right\rangle \left|\varepsilon\right\rangle \left|\varepsilon\right\rangle
\notag	\\
 = &
 \frac 1 4
\big[
\left|00\right\rangle \left(\left|0\right\rangle \left| \epsilon_{00} \right\rangle \left|0\right\rangle \left| \epsilon_{00} \right\rangle \right) 
\left|0000\right\rangle
+
\left|00\right\rangle \left(\left|0\right\rangle \left| \epsilon_{00} \right\rangle \left|1\right\rangle \left| \epsilon_{11} \right\rangle \right) 
\left|0001\right\rangle
\notag	\\
&+
\left|00\right\rangle \left(\left|1\right\rangle \left| \epsilon_{11} \right\rangle \left|0\right\rangle \left| \epsilon_{00} \right\rangle \right) 
\left|0010\right\rangle
+
\left|00\right\rangle \left(\left|1\right\rangle \left| \epsilon_{11} \right\rangle \left|1\right\rangle \left| \epsilon_{11} \right\rangle \right) 
\left|0011\right\rangle
\notag	\\
&+
\left|01\right\rangle \left(\left|0\right\rangle \left| \epsilon_{00} \right\rangle \left|0\right\rangle \left| \epsilon_{00} \right\rangle \right) 
\left|0100\right\rangle
+
\left|01\right\rangle \left(\left|0\right\rangle \left| \epsilon_{00} \right\rangle \left|1\right\rangle \left| \epsilon_{11} \right\rangle \right) 
\left|0101\right\rangle
\notag	\\
&+
\left|01\right\rangle \left(\left|1\right\rangle \left| \epsilon_{11} \right\rangle \left|0\right\rangle \left| \epsilon_{00} \right\rangle \right) 
\left|0110\right\rangle
+
\left|01\right\rangle \left(\left|1\right\rangle \left| \epsilon_{11} \right\rangle \left|1\right\rangle \left| \epsilon_{11} \right\rangle \right) 
\left|0111\right\rangle
\notag	\\
&+
\left|10\right\rangle \left(\left|0\right\rangle \left| \epsilon_{00} \right\rangle \left|0\right\rangle \left| \epsilon_{00} \right\rangle \right) 
\left|1000\right\rangle
+
\left|10\right\rangle \left(\left|0\right\rangle \left| \epsilon_{00} \right\rangle \left|1\right\rangle \left| \epsilon_{11} \right\rangle \right) 
\left|1001\right\rangle
\notag	\\
&+
\left|10\right\rangle \left(\left|1\right\rangle \left| \epsilon_{11} \right\rangle \left|0\right\rangle \left| \epsilon_{00} \right\rangle \right) 
\left|1010\right\rangle
+
\left|10\right\rangle \left(\left|1\right\rangle \left| \epsilon_{11} \right\rangle \left|1\right\rangle \left| \epsilon_{11} \right\rangle \right) 
\left|1011\right\rangle
\notag	\\
&+
\left|11\right\rangle \left(\left|0\right\rangle \left| \epsilon_{00} \right\rangle \left|0\right\rangle \left| \epsilon_{00} \right\rangle \right) 
\left|1100\right\rangle
+
\left|11\right\rangle \left(\left|0\right\rangle \left| \epsilon_{00} \right\rangle \left|1\right\rangle \left| \epsilon_{11} \right\rangle \right) 
\left|1101\right\rangle
\notag	\\
&+
\left|11\right\rangle \left(\left|1\right\rangle \left| \epsilon_{11} \right\rangle \left|0\right\rangle \left| \epsilon_{00} \right\rangle \right) 
\left|1110\right\rangle
+
\left|11\right\rangle \left(\left|1\right\rangle \left| \epsilon_{00} \right\rangle \left|1\right\rangle \left| \epsilon_{11} \right\rangle \right) 
\left|1111\right\rangle
\big]
\notag	\\
= 	& \frac 1 4
( \left|00000000\right\rangle+\left|00010001\right\rangle +\left|00100010\right\rangle + \left|00110011\right\rangle  
\notag	\\
&\phantom{(}+
\left|01000100\right\rangle + \left|01010101\right\rangle + \left|01100110\right\rangle + \left|01110111\right\rangle 
\notag	\\
&\phantom{(}+
\left|10001000\right\rangle + \left|10011001\right\rangle + \left|10101010\right\rangle +\left|10111011\right\rangle 
\notag	\\
&\phantom{(}+
\left|11001100\right\rangle + \left|11011101\right\rangle + \left|11101110\right\rangle + \left|11111111\right\rangle ) 
\left| \epsilon_{00} \right\rangle \left| \epsilon_{00} \right\rangle
\notag	\\
=& \left|\varUpsilon\left(p^1,p^2,\dots,p^8\right)\right\rangle 	\left| \epsilon_{00} \right\rangle	\left| \epsilon_{00} \right\rangle.
\end{align}
As can be seen from the above formula, no error can be introduced in the eavesdropping checking iff
the intercepted particles and ancillary particles are in a product state, 
hence the entanglement-measurement attack attack will not succeed.

\subsubsection{Trojan horse attack}

Trojan horse attacks refer to the delay-photon Trojan horse attack and invisible photon Trojan horse attack,
under which two-way quantum cryptography protocols have been proved to be insecure \cite{JiZX6562016}.
In the proposed protocol, all particles are transmitted to Alice and Bob by TP, and these particles do not return.
Therefore, the protocol is a one-way protocol, such that the Trojan horse attacks are naturally invalid.

\subsubsection{Photon-number-splitting attack}

The defense method against this attack is shown in the QPC protocol proposed by Chen et al. \cite{ChenXB28372010}. 
Specifically, the particles with illegal wavelengths can be filtered out by inserting filters
in front of the quantum devices of Alice and Bob.
Before performing measurements on sampling signals, Alice and Bob
use some beam splitters to split them for checking eavesdroppers.
If the multi-particle rate exceeds a predetermined threshold,
Alice and Bob terminate the protocol and restart it, otherwise they continue to carry out the protocol.

\subsubsection{Man-in-middle attack}

The specific means of this attack against the protocol adopted by Eve is as follows:
In the steps 2 and 3, Eve intercepts and captures the particles TP transmits to Alice.
Subsequently, she disguises herself as Alice and accomplishes eavesdropping checking process with TP.
Then, Eve performs measurements on the particles in $S_a$ and prepares fake decoy photons
and eight-qubit entangled states to replace $S_a^*$.
Finally, she transmits the fake photons and states to Alice.
Likewise, Eve disguises herself as TP to check whether eavesdroppers exist with Alice.
Similar to the intercept-measurement attack, using man-in-middle attack Eve can get $M_a^i$ and $M_b^i$,
thus she can only get $G_a^i \oplus G_b^i$ but not $G_a^i$ and $G_b^i$ by decryption.
In addition, Eve does not know the keys generated by QKD,
hence her attack will not succeed when Alice and Bob are in different locations.

Quantum authentication technology \cite{ZhangHG162019} can be used to resist the man-in-middle attack. 
When Alice and Bob are in different locations, if they use quantum authentication technology, 
Alice and Bob only need to employ QKD to create the keys $K_{ab}^i$.
Then, in the fourth step of the protocol, they calculate 
$G_a^i \oplus M_a^i \oplus K_{ab}^i$ and $G_b^i \oplus M_b^i \oplus K_{ab}^i $, respectively,
and use the symbols $R_a^i$ and $R_b^i$ to mark the computing results.
Finally, they jointly calculate $R_a^i \oplus R_b^i$ and publicly announce the computing result $ R_{AB}^i $ to TP.
After these changes, we still have
\begin{align}
(G_a^i \oplus M_a^i \oplus K_{ab}^i) \oplus (G_b^i \oplus M_b^i \oplus K_{ab}^i) \oplus M_{c_1}^i \oplus M_{c_2}^i
= G_a^i \oplus G_b^i,
\end{align}
thus the protocol output is still correct.

\subsection{Internal attack}

Two cases of the internal attack will be considered if the executors of the proposed protocol are curious. 
The first is that one participant is curious about another participant's data and wants to steal it.
The second is that TP is curious and tries to obtain the data of one or two participants.

\subsubsection{A curious participant's attack}

Without losing generality, suppose that Alice is curious about Bob's data and wants to steal it.
When they are in the same location, the only attack that Alice can take is to deduce
Bob's measurement results based on her measurement outcomes in the fourth protocol step.
For the eight-qubit entangled state showed in Eq. \ref{information-carrier}, 
we can calculate all the marginal states of $\left|\varUpsilon\left(p^1,p^2,\dots,p^8\right)\right\rangle$,
\begin{align}
\rho_{p^k} = Tr_{\{p^1,p^2,\dots,p^8\} \setminus \{p^k\}} 
\left(\left|\varUpsilon\left(p^1,p^2,\dots,p^8\right)\right\rangle \left\langle\varUpsilon\left(p^1,p^2,\dots,p^8\right)\right\vert \right)
= \frac I 2,
\end{align}
where $k=1,2,\dots,8$.
Therefore, Alice cannot get any useful information from her measurement results.

When Alice and Bob are in different places, due to no particles are exchanged between them,
if Alice wants to steal Bob's data, she can use Eve's attack methods mentioned above, 
including the man-in-middle attack, measurement-resend attack, intercept-resend attack, 
and entanglement-measurement attack. As analyzed above, these attacks are invalid.
Taking the man-in-middle attack as an example, Alice can get Bob's measurement result $M_b^i$ by this attack,
but she cannot steal Bob's data because she does not know the keys $K_{bc}^i$.
Of course, if the quantum authentication technology is used, 
this attack is invalid, and only the keys $K_{ab}^i$ is needed instead of the keys $K_{ac}^i$ and $K_{bc}^i$, 
which has been shown above.

\subsubsection{TP's attack}

As analyzed before, throughout the proposed protocol, the only information associated with the data of Alice and Bob
obtained by TP is either $G_A^i \oplus G_B^i$ or the data encrypted by $K_{ab}^i$. 
Therefore, TP cannot obtain Alice and Bob's data.


\section{Two variants}
\label{variants}

In this section, we will first show that replacing the single-particle measurements used in the proposed protocol 
with Bell measurements can also achieve the purpose of the protocol.
Then we will show that the proposed protocol can also be transformed into 
a semi-quantum private comparison protocol (SQPC) through several small changes.

\subsection{The protocol with Bell measurements}

The form of the eight-qubit entangled state $\left|\varUpsilon\left(p^1,p^2,\dots,p^8\right)\right\rangle$ can be 
rewritten as \cite{JiangS6052021}
\begin{align}
& \left|\varUpsilon\left(p^1,p^2,\dots,p^8\right)\right\rangle	\notag \\
=& \frac 1 4 
\left(
\left|\phi^+\right\rangle\left|\phi^+\right\rangle+\left|\phi^-\right\rangle\left|\phi^-\right\rangle 
+\left|\psi^+\right\rangle\left|\psi^+\right\rangle + \left|\psi^-\right\rangle\left|\psi^-\right\rangle 
\right)_{1234}
\otimes
\left(
\left|\phi^+\right\rangle\left|\phi^+\right\rangle+\left|\phi^-\right\rangle\left|\phi^-\right\rangle 
+\left|\psi^+\right\rangle\left|\psi^+\right\rangle + \left|\psi^-\right\rangle\left|\psi^-\right\rangle	
\right)_{5678},
\end{align}
where are Bell states (see Eq. \ref{Bell-sate}).

This variant requires the following minor changes to the original protocol:
\begin{enumerate}

\item 
TP, Alice and Bob agree on the two public coding rules:
$\left|\phi^+\right\rangle \leftrightarrow 00$, $\left|\phi^-\right\rangle \leftrightarrow 01$,
$\left|\psi^+\right\rangle \leftrightarrow 10$, and $\left|\psi^-\right\rangle \leftrightarrow 11$.

\item
Replace all the single-particle measurements used in the original protocol with Bell measurements.
TP, Alice and Bob mark the measurement results according to the new coding rules.

\end{enumerate}
It is easy to verify that after making the above changes, one can still get
\begin{align}
\label{output}
R_{AB}^i \oplus M_{c_1}^i \oplus M_{c_2}^i = G_a^i \oplus G_b^i.
\end{align}
That is, the output of the variant is correct.

\subsection{Mediated semi-quantum private comparison}

The concept of semi-quantum cryptography was first introduced by Boyer et al. by proposing 
a semi-quantum key distribution (SQKD) protocol in order to answer the question 
``how quantum must a protocol be to gain an advantage over its classical counterpart'' \cite{BoyerM99142007}, 
and then extended to other topics such as quantum quantum key agreement, quantum secure direct communication,
quantum secret sharing and quantum authentication
\cite{ZhangHG162019,IqbalH1932020}.

In a semi-quantum cryptography protocol, there is at least one fully quantum party with full quantum ability 
and at least one semi-quantum party with limited quantum ability 
(also called a classical party) \cite{ZhangHG162019,IqbalH1932020}.
Specifically, the fully quantum party has the ability to perform all quantum operations,
such as preparing and measuring any quantum state,
while the classical party can only perform some or all of the following operations \cite{ZhangHG162019,IqbalH1932020}:

\begin{enumerate}

\item Prepare qubits in the $Z$ basis.

\item Measure qubits in the $Z$ basis.

\item Measure the qubits from a sender in the $Z$ basis, and 
send her new qubits as measurement results.

\item Reflect the qubits back to a sender directly.

\item Reorder the qubits from a sender without disturbing them.

\end{enumerate}

We now point out that the protocol proposed in this paper and the protocol in Ref. \cite{JiZX34282019}
can be transformed into a mediated SQPC protocol through some slight changes,
where Alice and Bob act as the classical parties, while TP acts as the fully quantum party.
The concept of mediated SQPC comes from mediated SQKD
proposed by Krawec et al. \cite{KrawecWO9132015},
both of which introduce a fully quantum third party to help two classical parties Alice and Bob achieve their goals.
In the eavesdropping checking of the proposed protocol, Alice and Bob either perform measurements in the $Z$ basis or
reflect the qubits back to TP directly after receiving decoy photons. 
Then, TP, Alice and Bob can determine whether there is an eavesdropper in quantum channels
by comparing the initial states of the decoy photons with the ones after their operations 
(see Refs. \cite{BoyerM99142007,BoyerM7932009} for more details on eavesdropping checking).
For the keys $K_{ac}^i$ and $K_{bc}^i$ adopted in the proposed protocol, 
TP, Alice and Bob can generate them by using a SQKD protocol 
instead of a QKD protocol, while the keys $K_{ab}^i$ can be generated through a mediated SQKD protocol 
(e.g., Krawec et al.'s protocol \cite{KrawecWO9132015}).

We have shown that through the above changes, the proposed protocol and the protocol in 
Ref. \cite{JiZX34282019} can be transformed into a mediated semi-quantum protocol.
In the same way, the protocol in Ref. \cite{JiZX6562016} can also be transformed into a semi-quantum protocol.
Compared with quantum cryptography protocols, semi-quantum cryptography protocols require participants 
to have asymmetric quantum capabilities, which makes semi-quantum protocols possible 
to find more applications in real life.

\section{Conclusion}

The QPC protocol using eight-qubit entangled states has been presented.
We have considered the cases and protocol security when Alice and Bob are in the same and different places.
The proposed protocol utilizes the entanglement correlation of the eight-qubit entangled state,
joint computation between Alice and Bob, and keys generated by QKD to ensure protocol security,
which makes the external attack and internal attack invalid.
We have shown that the comparison of the secret data of Alice and Bob
can also be safely realized through Bell measurements. 
Since the proposed protocol only adopts single-particle measurements, 
it has been shown that the proposed protocol can be transformed into a semi-quantum protocol.
In the future, we will continue to focus on the usefulness of multi-qubit entangled states in QPC
and other branches of QSMC.

\section*{Acknowledgments}

This work is supported by National Key Research and Development Program of China (No.2020YFB2009502),
Fuzhou Science and Technology Project (No. 2018N2002 and No. 2019G50),
National Science Foundation of China (No. 31772045 and No. 31471704).


\end{document}